\documentclass[twocolumn,showpacs,preprintnumbers,amsmath,amssymb]{revtex4}

\usepackage{graphicx}
\usepackage{dcolumn}
\usepackage{bm}
\usepackage{caption}
\begin{document}

\preprint{APS/xxx}

\title{Localizations on Complex Networks\\}
\author{Guimei Zhu$^{1,2}$}
\author{Huijie Yang$^{2,3}$}
\author{Chuanyang Yin$^{1,2}$}
\author{Baowen Li$^{2,4}$}
\email{phylibw@nus.edu.sg} 
\address{$^1$ Department of Modern Physics,
         University of Science and Technology of China,
         Hefei Anhui 230026,
         China\\
         $^2$ Department of Physics and Centre for Computational Science
         and Engineering, National University of Singapore,Singapore
         117542\\
         $^3$ School of Management, Shanghai University for Science
         and Technology, Shanghai 200092, China\\
         $^4$ NUS Graduate School for Integrative Sciences and Engineering,
         Singapore 117597, Republic of Singapore}
\date{\today}

\begin{abstract}
We study the structure
characteristics of complex networks from the
representative eigenvectors of the adjacent matrix.
The probability distribution function
of the components of the representative eigenvectors are proposed to
describe the localizations on networks where the Euclidean distance
is invalid. Several quantities are used to describe the localization
properties of the representative states, as the participation ratio,
the structural entropy, the probability distribution function of the
nearest neighbor level spacings for spectra of complex networks. The whole cellular networks in real world, the
Watts-Strogatz small world and Barabasi-Albert scale-free networks are considered.
The networks have nontrivial localization properties due to the
nontrivial topological structures. It is found that the
ascend-ranked series of the occurring probabilities at the nodes
behave generally multi-fractal. This characteristic can be used as a
structure measure of complex networks.
\end{abstract}

\maketitle

\section{Introduction}
\label{intro}

Recent years have witness an avalanche of research on complex
network and its applications in diverse fields \cite{review}.
Structure measures of complex networks are the cornerstone to
understand relations between the structures, dynamics and functions.
Real world networks generally have nontrivial properties as the
small-world \cite{WattStrog98}, scale-free \cite{Barab99}, motif
\cite{Milo02}, modularity, hierarchy \cite{Rav02}, fractal
\cite{Song05} and so on. The small-world effect is that in average
the nodes can reach each other with only a small number of hops. The
scale-free refers to the number of edges per node obeys a
right-skewed distribution. It is also found that some special
subgraphs containing several connected nodes, called motifs, occur
with significant probabilities compared with that in the
corresponding randomized networks. These three individual, pair or
local pattern-based properties are called micro-properties. On the
other hand, the modularity is a kind of macro-property represents
that a network can be separated into loosely connected groups within
which the nodes are densely connected, respectively.

To a certain degree, dynamics on networks can be regarded as the
transport processes of mass, energy, signal and/or information at
different structure scales \cite{Song07,yang07}. Sometimes we have
to deal with networks with unreasonable large number of nodes and
edges, e.g., the neuron networks and the World-Wide-Web networks, when
designing a coarse-grain procedure is essential
\cite{kim04}. The patterns at different scales may provide a
reasonable solution to these problems. It is found that some real
world networks have hierarchical structures, in which the
small-world and the scale-free properties can coexist \cite{Rav02}.
Moreover, many real world networks behave self-similar at
different structure levels (fractal) \cite{Song05} .

Though great progresses have been archived, the measures of complex
networks are not yet fully understood. Just as pointed out by Newman \cite{newman03}, that our techniques for analyzing networks
are at present time no more than a grab-bag of miscellaneous and
largely unrelated tools, and we still do not have a systematic
program for characterizing network structures. Furthermore, the
measures of network structures, such as the micro-properties, the
patterns at different scales and the macro-properties, are generally
a simple application of the concepts in graph theory,
bioinformatics, social science and fractal theory, namely, they are
not dynamics-based. We can not expect simple and reasonable
relations between the measures and the dynamical processes on
networks.

The lack of powerful tools to characterize network structures is an
essential bottleneck to understand dynamical processes on
networks. One typical example is the synchronizabilities of complex
networks. Detailed works show that almost all the structure measures
affect the synchronizabilities in complicated ways \cite{zhoucs06},
based upon which we can not reach a clear picture of the mechanisms
for synchronization processes on networks.

Dynamics-based measures of complex networks may be the key to the
problems. The structures of complex networks can induce nontrivial
properties to the physical processes occurring on them.  The
physical processes in turn can be used as the probe to capture the
structure properties. Well studied dynamical processes, such as the
random walks \cite{sood07,costa07} and the Boolean dynamics
\cite{boolean}, can be good candidates as probes. To cite an
example, the random walks on complex networks that biased towards a
target node show a localization-delocalization transition
\cite{sood07}.

In the present paper, we map networks to large clusters, namely, the
nodes and edges to atoms and bonds between them, respectively. The
localization properties of electrons in the clusters can be used as
measures of the structure properties of the networks. We try to detect the global
symmetries from the spectra and the eigenvectors of complex
networks. Very recently, much attentions have been focused on
detecting global characteristics embedded in spectra of complex
networks due to their potential application in understanding the
organization mechanisms and the synchronization dynamics of complex
networks \cite{spec4,yang04,yang07}. To our best knowledge, it is
the first time to detect the global characteristics of complex
networks from the eigenvectors which contain more information about the system than eigenvalues.

Besides as a measure of network structures, the structure-induced
localization may have potential application in understanding the
electronic properties of materials such as conductive polymers and carbon
nanonets. The intrachain windings in conductive polymers can
introduce long-range edges into the original one dimensional (1-D)
systems, resulting in nontrivial network structures
\cite{Xiong95,kim2007}. It is also found that random networks of
carbon nanotubes, called nanonets, can mimic a variety of basic
electronic functions from the conductive properties of metals to the
less conductive characteristics of semiconductors \cite{Nanonets}.
Indeed, nanonets have paved the way for the carbon to serve as the
foundation for future electronic devices. The effect of network structures
on electronic properties is one of the most active topics in recent
years \cite{Song07,Stanley}.

The paper is organized as follows: In Sec.II the concept of
localization on complex networks is presented. The occurring
probabilities on the nodes are proposed to describe quantitatively
the localization effects. In Sec. III the methods to measure the
localization properties are described in detail. The participation
ratio, the structural entropy and the probability distribution
function of the nearest neighbor level spacings of spectra are used
to illustrate the localization in a global way. The wavelet
transform is then used to find the detailed structure properties of
the probability distribution function of the occurring probabilities
on the nodes. As examples, we consider the Watts Strogatz
small-world and the Barabasi-Albert scale-free modeling networks and
the whole cellular networks in real world. The results are shown in
Sec. IV.  We will be shown that the global symmetries in networks can
induce multi-fractal structures in the eigenvectors. As a
conclusion, the nontrivial structures of complex networks can induce
significant localizations, which in turn can be used as a global
measure of the structure symmetries.

\section{The Localizations on Networks}
\label{sec:2}

We consider a undirected complex network with $N$ identical nodes,
whose topological structure can be described by an
adjacent matrix $A$. The elements $A_{ij}$ are $1$/$0$ if the
nodes $i$ and $j$ are connected/disconnected, respectively. If
we consider the nodes as atoms and the edges as bonds, the network
can be mapped to a large molecule \cite{yang04}. For an electron
moving in such a molecule, the tight-binding Hamiltonian is,

\begin{equation}
\label{eq1} \mathcal H = \sum\limits_{n = 1}^N {\varepsilon _n \cdot
\left| n \right\rangle \left\langle n \right|} + \sum\limits_{m \ne
n}^N {A_{mn}\cdot t_{mn} \cdot \left| m \right\rangle \left\langle n
\right|} ,
\end{equation}

\noindent where $\varepsilon _n$ is the site energy and $t_{mn}\cdot
A_{mn}$ the hoping integral for the bond between sites $m$ and $n$.

A tight-binding Hamiltonian of Eq.(1) is usually used to study the
disorder-induced localizations. In the present form, the matrix $A$
is explicitly introduced to describe the structure of the system.
For a one dimensional (1-D) perfect regular lattice, we have
$\varepsilon _n=\varepsilon, t_{mn}=t$ and $A_{mn}=\delta (m-n \pm
1)$. The Bloch wave function of an electron extends all over this
perfect regular lattice. Disorder structures can induce a transition
from extended to localized states. The wave function for a localized
state decreases exponentially with the distance from its center. The
disorder effects include the random distributions of the site
energies ($\varepsilon _n$), the hoping integrals ($t_{mn}$) and the
edges in structures ($A_{mn}$). The disorders come from the
different kinds of atoms on the lattice points, the differences of
the separations of successive lattice points and the randomness in
structures. At the same time, there may be some symmetries in the
distributions of the site energies, the hoping integrals and the
edges, which may lead to delocalization of the wave functions.

In the usual Anderson model \cite{ander}, the disorder effect due to
the random distribution of the site energies is considered, i.e.,
$\varepsilon _n$ is a random variable satisfying a certain
distribution probability function while $t_{mn}=t,A_{mn}=\delta (m-n
\pm 1)$. The site energies may obey a special distribution rather
than that in the Anderson model, as a periodic \cite{kim89} or a
power-law \cite{titov03} function. In literatures
\cite{tang86,tang87}, a one-dimensional quasicrystal model is
introduced that the separation of two successive lattice points
takes one of the two values $u$ and $v$. This model considers the
disorder effect of the distribution of the hoping integrals. We have
$\varepsilon _n$=const., $t_{mn}=t_u$ or $t_v$ and $A_{mn}=\delta
(m-n \pm 1)$. $t_u$ and $t_v$ are the hoping integrals corresponding
to the separations $u$ and $v$, respectively. It is found that
quantum systems with quasi-periodic structures will be in an
intermediate state, which can be described with critical wave
functions. A critical wave function obeys a power-law with respect
to the distance from its center.

To investigate the problems as vibration spectra of glasses ,
instantaneous normal modes in liquids, electron hopping in amorphous
semiconductors and combinatorial optimization, Euclidean random
matrix (ERM) models are widely used in literatures \cite{erm}, in
which the disorder is due to the random positions of the sites, and
the matrix elements are given by a deterministic function of the
distances.

The models mentioned above generally focus on the disorder effects
of the site energies and the hopping integrals. These models have
also been extended to nontrivial structured systems such as the Cayley
tree \cite{cayley} and the small-world networks \cite{Xiong95}.
Nontrivial effects of the structures of the systems are reported,
but the interplay between the disorders due to the site energies and
that due to the structures makes it difficult to distinguish the
structure disorder effect from the site energy disorder effect.

In the networks considered in the current paper, however, the nodes are
all identical and the disorder effect comes from the nontrivial topological structure. We focus our attentions on the disorder
effect of the network structure, that is, we assign $\varepsilon
_n=0$ and $t_{mn}=1$, which leads $H = A$. The localization on the
network refers to the network structure-induced characteristics of
the wave functions for this system. The usual Anderson model
\cite{ander} is a special case that in the networks there exist
connections only between the nearest neighbors in Euclidean space.

Statistically, the structures of networks should display certain
symmetries due to the general rules obeyed in the construction of
the networks. Recent works demonstrate that many theoretical and
real world networks have statistically self-similar structures
\cite{Song05}. Therefore, there are two competitive mechanisms
determining the wave function property, the randomness of the bonds
in the netowkrs tends to cause localization of wavefunction, whereas
the symmetries of the networks intend to make wave function
extended. We thus expect
 rich structures embedded in the wave
functions. As it is well known, aperiodic crystals lead to the
fractal wave functions \cite{tang87,tang87b}. An interesting
question is then, how the global symmetries of networks affect the
localization properties. The localization can be used as a probe of
the characteristics of the network structures.

The states in the center of the energy band have the best chance to
remain as extended for a moderately disordered system. The
eigenvector corresponding to the special eigenvalue close to the
center of the spectrum for a network, denoted by $E_c $, is employed
as the representative state to illustrate the characteristics of the
considered system.

In the traditional study of  wave function localization,
the physical systems have deterministic structures in real world
Euclidean space, which leads to natural definitions of the localized, intermediate and extended states of the systems.
Obviously, these definitions are invalid for general complex
networks without deterministic structures in Euclidean space. In
this paper, we describe the localization effects with the
probability distribution function (PDF) of the occurring
probabilities at the nodes, i.e., the values of the components for
the representative eigenvector. Based on the PDF of the occurring
probabilities, the traditional definitions are extended to a much
more general version to describe the localization properties on
complex networks.

In Euclidean space, for a state $\Psi(r)$, the occurring probability
is $\rho(r)=\left| {\Psi (r)} \right|^2 \equiv F(r)$. Because the
value of the distance $r$ distributes homogeneously in the
considered region, we can regard it as a homogenously distributed
random variable. The direct sampling method in Monte Carlo
simulations tells us that the probability distribution of $\rho$
should be $P(\rho) \propto \frac{dF^{ - 1}(\rho )}{d\rho }$. Hence,
it is reasonable to define the localized, critical and
perfectly extended states on complex networks with the PDFs of the
occurring probabilities,

\begin{equation}
\label{eq1} P(\rho ) \propto \left\{ {{\begin{array}{*{20}c}
 {\delta (\rho - \rho _0 ),} \hfill & {extended} \hfill \\
 {\rho ^{ - (1 + \eta )}\left| {\eta > 0} \right.,} \hfill & {critical}
\hfill \\
 {\rho ^{ - (1 + \eta )}\left| {\eta = 0} \right.,} \hfill & {localized}
\hfill \\
\end{array} }} \right..
\end{equation}

\noindent The PDF of the representative function is a very
powerful measure to capture the localization properties. It can be
used to find the localization properties without using
distance in real world Euclidean space.

Because no derivative exists for a fractal wave-function in
Euclidean space, the extension procedure in defining critical and
localized states on networks can not be simply used to define
fractal property on networks with the PDF of the occurring
probabilities. In the present paper, we detect directly the fractal
characteristics in the ascend-ranked series of the the occurring
probabilities, as described in detail in Section $III(C)$.

\section{Methods}
\label{sec:3} \subsection{Structural Entropy}

We denote the representative state with $V = (V_1 ,V_2 , \cdots ,V_N
)$. The occurring probabilities at the nodes are $\rho_m=\left|
{V_m} \right|^2,m=1,2,\cdots,N$. The localization extent of the
state can be described quantitatively with the participation ratio
\cite{Janssen,Mirlin},

\begin{equation}
\label{eq2} Q = \frac{1}{N \cdot \sum\nolimits_{m = 1}^N {\rho
_m}^2}.
\end{equation}

\noindent For a perfect extended state we have $Q =1$, while for a
state strongly localized on one node it tends to $\frac{1}{N}$.
Generally, $Q$ should be in the range of $\left[ {\frac{1}{N},1}
\right]$.

However, this participation ratio can capture only the primary-level
complexity in the localization properties, namely, the extension of
the representative eigenvector to $N \cdot Q $ nodes on the network.
Many PDFs with different localization behaviors may result in the
same $Q$. The simplest one is a step-like function that on $N \cdot
Q$ nodes the occurring probabilities are $\frac{1}{N \cdot Q}$,
while on the left $N \cdot (1-Q)$ nodes the occurring probabilities
are $0$.

The secondary-level complexity in the localization properties is the
deviation of the PDF from the step-like function. This deviation
corresponds to the shape of the PDF, which can be extracted by using
the structural entropy \cite{Pipek92},

\begin{equation}
\label{eq3} S_{str} = -\sum\limits_{m = 1}^N {\rho
_m}\ln{\rho_m}-\ln(Q \cdot N).
\end{equation}

\noindent For the simple step-like condition, we have $S_{str}=0$.
$S_{str} \ne 0$ tells us the shape deviation of PDF from the simple
one.

The pair of localization quantities, $(Q,S_{str})$, is widely used
up to date to describe the localization
in disordered and aperiodic systems, and the statistical analysis of
spectra in diverse fields such as quantum chemistry, condensed matter physics, and quantum chaos \cite{Pipek03,Varga02}.

\subsection{Statistical Properties of the Spectra}

The localization property can also be described with the random
matrix theory (RMT) \cite{guhr,dittrich,Mirlin}. RMT is initially
developed to understand the energy levels of complex nuclei and
other kinds of complex quantum systems. Recently, the RMT theory has
been proposed to capture the structure and dynamical properties of
complex networks \cite{spec4}.

One of the most important quantities in the theory is the PDF for the
nearest neighbor level spacings (NNLS) of the spectrum. It is theoretically and numerically confirmed
that at the localization and the extended states the PDFs of the
NNLS should be Poisson and Wigner-Dyson distribution, respectively
\cite{Altshuler,Shklovskii,Hofste}. Generally, for an intermediate
state, the PDF obeys the Brody distribution,

\begin{equation}
\label{eq4} U(s) = \frac{\beta}{\xi} s^{\beta-1}exp{\left[
  -\left(\frac{s}{\xi}\right)^\beta\right]},
\end{equation}

\noindent where $s$ is the NNLS and $\xi$ the characteristic
distribution width. The Poisson and the Wigner-Dyson distributions
are the two extremes with $\beta=1$ and $\beta=2$, respectively.

Introducing the accumulated function, $C(s)=\int_{0}^{s} U(x)dx$,
some trivial calculations lead to,

\begin{equation}
\label{eq5} \ln R(s) \equiv \ln \left[ {\ln \left( {\frac{1}{1 -
C(s)}} \right)} \right] = \beta \ln s - \beta \ln \xi.
\end{equation}
\noindent From this formula we can testfy the Brody distribution and determine reliably the values of the parameters $\beta$ and $\xi$.

To make the spacings $s$ in units of local mean
level spacing, we should conduct a standard procedure, called
unfolding. Denoting the spectrum of a network with
$\lambda_1,\lambda_2,\cdots,\lambda_N$, the accumulation density
function for the spectrum is $G(\lambda_m)=m,m=1,2,\cdots,N$.
Fitting this relation with a polynomial function, we can separate it
into a smooth part and a fluctuation part as,

\begin{equation}
\label{eq6} G(\lambda_m)=G_{av}(\lambda_m)+G_{f}(\lambda_m).
\end{equation}

\noindent The NNLS can be obtained as,
$s_i=G_{av}(\lambda_{i+1})-G_{av}(\lambda_i),i=1,2,\cdots,N-1$. For
a complex network, we generally have not enough knowledge on its
spectrum, when the polynomial function fitting method can lead to a
reliable result. In this paper, the order of the polynomial function
is $17$.

\subsection{Wavelet Transform}

The detailed properties for the PDF of the occurring probabilities
can be used as the measure of the global structure symmetries.
However, determining this PDF is a nontrivial task \cite{new07}.
Assume the probability values have been sorted in ascending order,
namely, $\rho=\{\rho_1 \le \rho_2 \cdots \le \rho_N\}$, which can be
regarded as the profile of the nearest spacing series,
$\Delta\rho=\{\rho_2-\rho_1, \rho_3-\rho_2, \cdots,
\rho_N-\rho_{N-1}\}$. The local structures of $\Delta\rho$ can tell
us the probability distribution function of $\rho$. It is found that
the series $\rho$ generally behaves multi-fractal.

The wavelet transform (WT) \cite{Ivanov99} is used to detect the
fractal properties embedded in the ascend-ranked series $\rho$. The
increasing trend in the series $\rho$ makes the box-counting-based
techniques invalid to quantify the local scalings. In the wavelet
transform, the contributions of the polynomial trends can be removed
effectively. A multi-fractal series can be decomposed into many
subsets characterized by different local Hurst exponent $h$, which
quantifies the local singular behavior and thus relate to the local
scaling of the series. The statistical properties of these subsets
can be quantified by the fractal dimension $D(h)$ of the subset
whose local Hurst exponent is $h$.

As a standard procedure, we first find the WT maximal values,
$\left\{ {T_g (a,\rho_k (a))\left| {k = k_1 ,k_2 , \cdots k_J }
\right.} \right\}$, where $a$ is the given scale. The partition
function should scales in the limit of small scales as,

\begin{equation}
\label{eq4} Z(a,q) = \sum\limits_{k = k_1 }^{k_J } {\left| {T_g
(a,\rho_k (a))} \right|} ^q\sim a^{\tau (q)}.
\end{equation}

\noindent The fractal dimension $D(h)$ can be obtained through the
Legendre transform,

\begin{equation}
\label{eq5}
\begin{array}{l}
          D(h) = qh - \tau (q), \qquad h = \frac{d\tau (q)}{dq}.\\
\end{array}
\end{equation}

\noindent For a mono-fractal structure we have a linear relation,
 $\tau (q) = qH - 1$. $H$ is the global Hurst exponent. For positive
and negative $q$, $\tau (q)$ reflects the scaling of the large
fluctuations and small fluctuations, respectively.

We use the real analytic wavelet $g^{(n)}$ among the class of
derivatives of the Gaussian function, by which the polynomial trends
up to $n$ order can be removed. The results with $n=7$ are
presented. $n=5$ and $n=6$ lead almost the same results. As
comparison, we detect also the scaling behaviors in the randomized
series $\rho_R$, called shuffled series.

In this paper, we are interested in the characteristic point at which the
fractal dimension reaches its maximum value, $(h_c,D(h_c))$. It can
tell us the non-homogeneous distribution of the series $\rho$ and
the fractal characteristics of the principal subset.

\section{NUMERICAL RESULTS}

We examine the localization behaviors for the cellular networks
\cite{cellular}, which are compiled by using a graph-theoretical
representation of all the biochemical pathways based upon the WIT
integrated-pathway genome database of $43$ species from Archaea,
Bacteria and Eukarya \cite{WIT}. The whole cellular networks
consider the cellular functions as intermediate metabolism and
bioenergetics, information pathways, electron transport, and
transmembrane transport. The direct edges are replaced simply with
non-directed edges.  We consider only the cellular networks with the
sizes larger than $500$.

\begin{figure}
\scalebox{0.85}[0.85]{\includegraphics{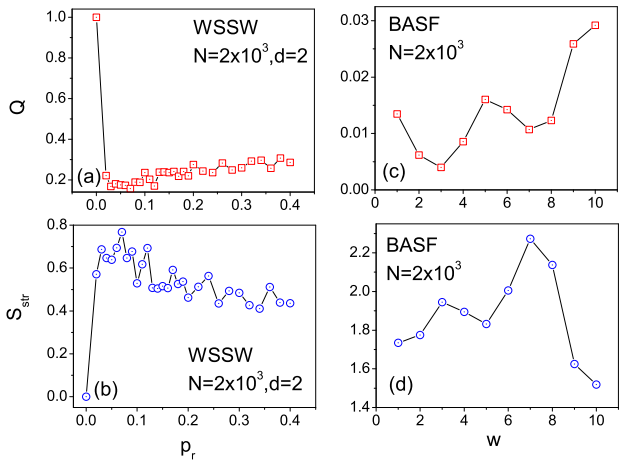}}
\caption{\label{fig:epsart}(Color online) The localization
quantities $(Q,S_{str})$ for the WSSW and BASF networks. There exist
complex relations between $p_r$ or $w$ and $(Q,S_{str})$ for the
BASF and WSSW networks. For the WSSW networks, from $p_r=0$ to
$p_r=0.02$ there exists an abrupt decrease/increase in value of $Q$/
$S_{str}$, as shown in (a)-(b)) respectively. Then with the increase
of the rewiring probability $p_r$ the participation ratio tends to
increase while the structural entropy tends to decrease; (c)-(d)
Results for the BASF networks.}
\end{figure}

\begin{figure}
\scalebox{0.85}[0.85]{\includegraphics{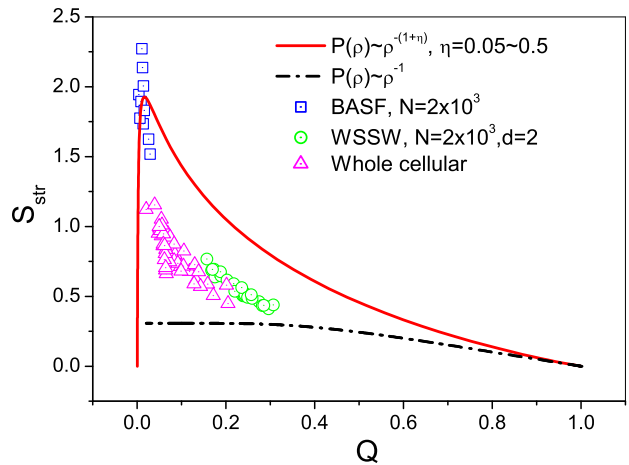}}
\caption{\label{fig:epsart} (Color online) The relations of
$S_{str}$ versus $Q$ for the WSSW, BASF and whole cellular networks.
The localization quantities for the distributions, $P(\rho) \sim
\rho^{-(1+\eta)}$ and $P(\rho) \sim \frac{1}{\rho}$, namely, the
critical and localized sates, are shown as references. Starting from
$\rho(r) \sim r^{-\sigma}$, the set of normalized values of
$\rho(\frac{n}{N}),n=1,2,\cdots,N$ can be regarded as the critical
state. Assigning $\sigma=1 \sim 10$, the corresponding values of
$\eta$ are $0.5 \sim 0.05$. $N$ is the size of the considered
networks. The same procedure is also used to generate the localized
states by starting from $\rho(r)\sim exp(-\mu r)$. The localized
states with $\mu=0.01 \sim 100 $ are generated. The localization
properties of the BASF networks can be captured by the critical
state with extremely significantly small values of $\eta$. The WSSW
and whole cellular networks are generally in between the two typical
states.}
\end{figure}

We study also the localization behaviors for the the
Watts-Strogatz small-world (WSSW) \cite{WattStrog98} and the
Barabasi-Albert scale-free (BASF) \cite{Barab99} networks. For the
WSSW model, we construct firstly a regular circle lattice with each
node connecting with its $d$ right-handed nearest neighbors. For
each edge we rewire it with probability $p_r$ to another randomly
selected node. Self- and double-edges are forbidden. By this way, we
can introduce randomness into the resulting networks. Moreover,
compared with that for the initial regular lattice, the rewiring
procedure may introduce also ``long-range" edges to the resulting
networks, which can reduce significantly the average number of hops
required for the nodes to reach each other. This is the so-called
small-world effect.

The BASF networks are the results of a preferential growth
mechanism, which exists widely in diverse fields. Starting from
several connected nodes as a seed, at each growth step a new node is
added and $w$ edges are established between this node and the
existing network. The probability for an existing node being
connected with the new node is proportional to its degree. Self- and
double-edges are forbidden. For the resulting networks, the number
of edges per node obeys a power-law, namely, no
characterized scale exists in this distribution.

\begin{figure}
\scalebox{0.85}[0.85]{\includegraphics{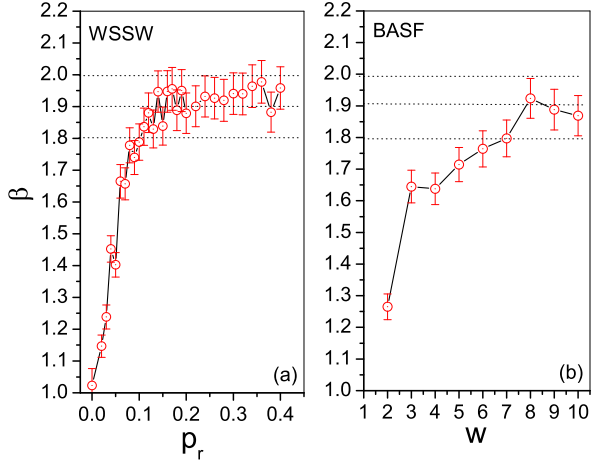}}
\caption{\label{fig:epsart} (Color online) The value of Brody
parameter $\beta$ versus network parameters $p_r$ and $w$. (a) WSSW
networks; (b) BASF networks.}
\end{figure}

\begin{figure}
\scalebox{0.85}[0.85]{\includegraphics{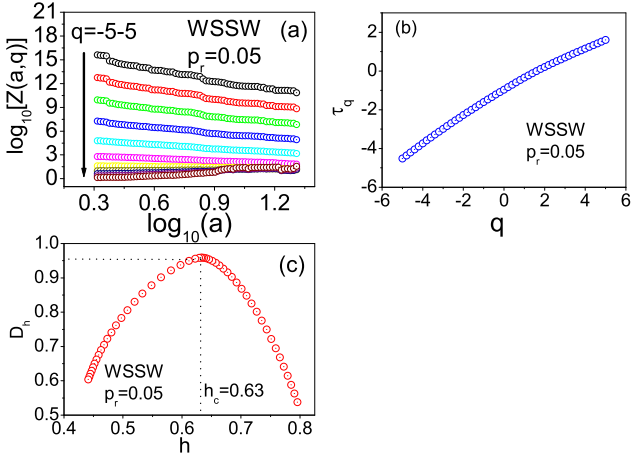}}
\caption{\label{fig:epsart} (Color online) The multi-fractal scaling
characteristics of the ascend-ranked series $\rho$ for the real
world, the WSSW and the BASF networks. The multi-fractal behavior
for the WSSW network with $d=2, N=2000$ and $p_r=0.05$ is presented
as a typical example. In the whole range of $q=-5\sim 5$, there is
only one characteristic point, $(h_c,D(h_c))=(0.63,0.958)$. }
\end{figure}

Figure 1 presents the localization quantities, $(Q,S_{str})$, for
the networks.  For the WSSW networks, the randomness introduced by
the rewiring procedure has two competitive effects, the long-range
edges which favors the extension, and the broken of symmetry which
induces the localization. For the BASF networks, the increase of $w$
increases the heterogeneity and the connections between the
nodes, which induce the localization and the extension,
respectively. Hence, there exist complex relations between $p_r$ or
$w$ and $(Q,S_{str})$ for the two kinds of networks, as shown in
Fig.1(a)-(b) and (c)-(d), respectively. For the WSSW networks, the participation ratio decreases rapidly from
$1$ to $0.22$ when $p_r$ changes slightly from
0 to 0.02, and then goes up gradually with the increase of $p_r$.
As for the structural entroy, it increases abruptly
when $p_r$ changes from 0 to 0.02, after that it decreases gradually with
the increase of $p_r$.

\begin{figure}
\scalebox{0.85}[0.85]{\includegraphics{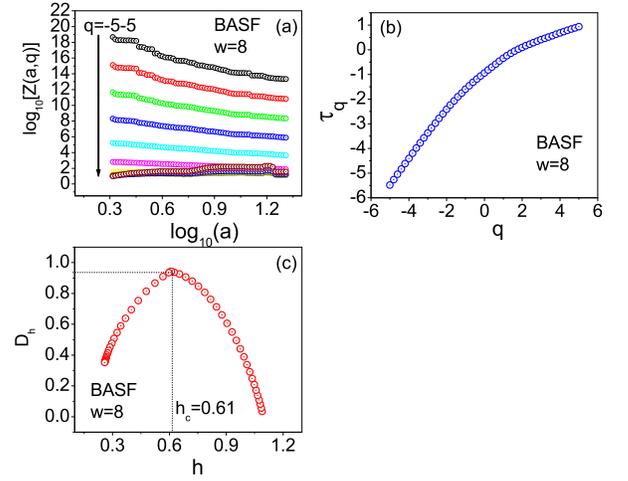}}
\caption{\label{fig:epsart} (Color online) The multi-fractal scaling
characteristics of the ascend-ranked series $\rho$ for the real
world, the WSSW and the BASF networks. The multi-fractal behavior
for the BASF network with $w=8$ and $N=2000$ is presented as a
typical example. In the whole range of $q=-5\sim 5$, there is only
one characteristic point, $(h_c,D(h_c))=(0.61,0.942)$.}
\end{figure}

\begin{figure}
\scalebox{0.85}[0.85]{\includegraphics{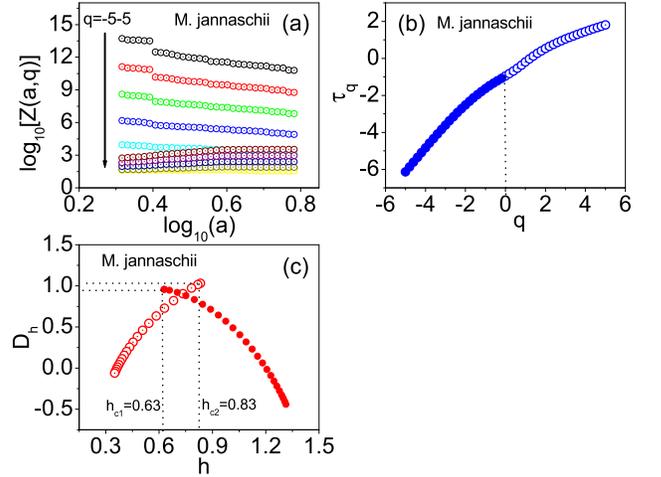}}
\caption{\label{fig:epsart} (Color online) The branched
multi-fractal scaling characteristics of the ascend-ranked series
$\rho$ for the real world and modeling networks. The branched
multi-fractal behavior for the whole cellular network of $M.
jannaschii$ is presented as a typical example. The two branches
$q<0$ and $q>0$ lead to different characteristic points, $(h_{c1},
D(h_{c1}))=(0.63,0.96)$ and $(h_{c2}, D(h_{c2}))=(0.83,1.03)$.}
\end{figure}

Figure 2 shows $S_{str}$ versus $Q$. As references,
we calculate also the localization quantities for the critical and
the localized states, namely, $P(\rho)\sim \rho^{-(1+\eta)}$ and
$P(\rho) \sim \frac{1}{\rho}$, respectively. Starting from $\rho(r)
\sim r^{-\sigma}$, we calculate the values of
$\rho(\frac{n}{N}),n=1,2,\cdots,N$. The resulting normalized values
can be regarded as the localized state. The critical sates with
$\sigma=1 \sim 10$ are calculated, and the corresponding values of
$\eta$ are $0.5 \sim 0.05$, respectively. $N$ is the size of the considered
networks. The same procedure can be used to generate the localized
states by starting from $\rho(r) \sim exp(-\mu r)$. The localized
states with $\mu=0.01 \sim 100$ are generated.

The localization properties of the BASF networks can be captured by
the critical states with extremely small values of $\eta$. The WSSW
and whole cellular networks are generally inbetween the two
typical (localized and extended) states .

We can find that the PDFs of the NNLS for all the networks can be
described very well by using the Brody distribution in a unified
way. The results for the parameter $\beta$ are shown in Fig.3. For
the WSSW networks, with the increase of the rewiring probability
$p_r$, the parameter $\beta$ increases rapidly from $1.02 \pm 0.053$
at $p_r=0$ to $1.95 \pm 0.065$ at $p_r=0.14$. For $p_r> 0.14$,
$\beta$ are almost same, namely $\sim 2.0$. That is, the
representative eigenvector changes from a nearly localized state
($p_r=0$) to an extended state in this interval of $p_r$. For the
networks with $p_r>0.14$, the representative eigenvectors are almost
perfectly extended. While for the BASF networks, with the increase
of $w$, the more edges can induce the significant extensions of the
representative states. $\beta$ reaches its asymptotic value $\sim
1.90$. Due to the heterogeneity, the BASF networks can not reach a
perfectly extended state.

In a considerable wide range of $q$, the partition functions behave
scale-invariant as in Eq.(8). There are three kinds of typical WT
transform results. Here we present several typical examples. In the
whole range of $q=-5 \sim 5$, the WSSW network with $p_r=0.05$ and
the BASF network with $w=8$ are multi-fractal with only one
characteristic point $(h_c,D(h_c)$, as shown in Fig.4 and Fig.5,
respectively. Sometimes, the multi-fractal degenerates to
mono-fractal. Fig.6 gives another condition where the fractal
behaviors can be separated into two branches, namely, $q<0$ and
$q>0$. The characteristic points for these two branches are not
same. That is, the principal subsets for the large fluctuations and
the small fluctuations are different. These three conditions are
called mono-fractal, multi-fractal and branched multi-fractal,
respectively.

The scaling properties for the real world networks and the modeling
networks are listed in Table I. For the mono- and multi-fractals, we
present simply the global Hurst exponent $H$ and the characteristic
point $(h_c,D(h_c))$, respectively. For the branched multi-fractal
we give the scaling characteristics for the two branches $q<0$ and
$q>0$, which are separated by the division symbol ${"/"}$. To cite
an example, for the cellular network $M. jannaschii$, the
characteristic point for the branch $q<0$ is  $(0.63,0.96)$ and that
for the branch $q>0$ is $(0.83, 1.03)$. It is denoted with
$(0.63,0.96)/(0.83,1.03)$. The results for the corresponding
shuffled series are presented also. We discard the networks that the
scaling behaviors of the original $\rho$ and the randomized series
$\rho_R$ are undistinguishable. The sizes of the WSSW and BASF
networks are $N=2000$. $N=1000,3000$ and $4000$ lead almost same
results (not shown in Table.1).

The WSSW and BASF networks are almost all mono- or multi-fractals
with the values of $h_c$ mainly in the range of $0.66 \pm 0.05$.
However, most of the considered real world networks behave branched
multi-fractal. The Hurst exponents lager than $1$ and near $0$
correspond to non-singularity and white noises, respectively.
Discarding these trivial conditions, we find that the
multi-fractal behaviors are embedded in the branches of $q>0$. And
the values of $h_c$ are basically in the range of $0.8\pm 0.05$. The
larger values of $h_c$ for the large fluctuations in the real world
networks show us the much more non-homogeneous structures of the PDF
of $\rho$. That is, the PDF of $\rho$ for the real world networks
tend to form much sharper peaks at different scales.

It should be noted that in the present paper the structure-induced
localization is used as a probe of structure properties of complex
networks. We detect the localization properties for the WSSW, BASF
model networks and the cellular networks, but it dose not imply and
require any localization-related dynamical process (such as waves)
occurring on the real-world systems.

\section{CONCLUSIONS}

In summary, the probability distribution function of the
representative eigenvector is proposed to describe the localization
properties of complex networks. The localization quantities
$(Q,S_{str})$, the PDFs of the NNLS and the wavelet transform are
used to capture the characteristics of the representative state. The
nontrivial structures of the networks can induce the localizations
of the representative states. At the same time, because of the
global symmetries of the networks, the representative state have
nontrivial structures rather than the step-like distribution.

The localization quantities $(Q,S_{str})$ and Brody distribution parameter $\beta$ can describe
the nontrivial localization properties in a global way. The
$(Q,S_{str})$ values tell us that the BASF networks with $w=2$ are
significantly localized compared with the WSSW networks with $d=2$.
It is consistent with the conclusions drawn from the results of
$\beta$. The whole cellular networks have localization properties
much closer with the WSSW networks.

The wavelet transform can tell us the details on the nontrivial
structures of the representative eigenvectors. The ascend-ranked
series $\rho$ for the WSSW and BASF modeling networks and the whole
cellular networks behave mono-fractal, multi-fractal or branched
multi-fractal. The PDF of $\rho$ tends to form sharp peaks at
different scales in a self-similar way.

This kind of property can shed light on the global symmetries due to
the general rules in the construction of the networks. Hence, it can
be employed as a global measure of the network structures.
Moreover, the structure-induced localizations on networks maybe helpful
to understand the electronic conduction and heat transport
properties \cite{liu2007} of nanonet materials.

A closely relevant topic is the diffusion on complex networks. Kim
et al. \cite{kim2003} report for the first time their works on
quantum and classical diffusion on WSSW networks. The Hamiltonian is
same as that in the present paper, namely $\varepsilon_n=0$ and
$t_{mn}=1$ in Eq.(1).  An electron is localized at a randomly
selected node at beginning, then the diffusion process is obtained
by solving the time-dependent Schrodinger equation. It is found that
the ''long-range" edges can fasten the diffusion speeds
significantly, especially at the transition point from $p_r=0$ to
$p_r \ne 0$. This is qualitatively in consistent with our findings
of the significant changes of the participation ratio and the
structural entropy, $(Q,S_{str})$, when $p_r$ increase from $p_r=0$
to $p_r=0.02$.

As for the classical diffusion on networks, a very recent work
reports the first-passage times (FPT) of random walkers in complex
scale-invariant media \cite{condamin07,costa07}. Many real-world
networks have self-similar structures, and diffusion on networks can
be regarded to a certain degree as the diffusion on fractal media,
which has attracted intensive attentions for its importance in
theories and potential use in diverse research fields
\cite{anh2007}.

However, we shoud point out that, it is not trivial to compare our results quantitatively with these
evolution processes. Actually our results are obtained from the
eigenstates in energy representation, while for the quantum
diffusion the initial state of localizing at a randomly selected
node is a wave packet and the final state should be a superposition
of the eigenstates in energy representation. How to relate the
localization with the classical diffusion is definitely interesting but not a trivial task.
Obviously, detailed works on diffusion on complex networks are
required to understand the relation between localization and
diffusion on networks.

\begin{center}
\textbf{Acknowledgements}
\end{center}

The work is supported by NUS Faculty Research Grant No.
R-144-000-165-112/101. It is also supported by the National Science
Foundation of China under Grant Nos.70571074 and 10635040. H. Yang
gratefully acknowledges the support of K.C.Wong Education
Foundation, Hong Kong.

\end{document}